\documentclass[aps,pra,reprint,superscriptaddress]{revtex4-2}
 
 \usepackage[english]{babel}
\usepackage[dvipsnames,svgnames,x11names]{xcolor}
\usepackage[markup=underlined]{changes}

\definechangesauthor[color=BrickRed]{DW}
\definechangesauthor[color=NavyBlue]{XXX}

\usepackage{amsmath}
\usepackage{amssymb}
\usepackage{amsfonts}
\usepackage{bbm}
\usepackage{bm}
\usepackage{graphicx}
\usepackage{color}
\usepackage{dcolumn}
\usepackage{MnSymbol}
\usepackage{times}
\usepackage[colorlinks=true,linkcolor=blue,citecolor=blue,urlcolor=blue]{hyperref}

\newcommand{\ket}[1]{|{#1}\rangle}

\usepackage{color}
\definecolor{dred}{rgb}{.8,0.2,.2}
\definecolor{ddred}{rgb}{.8,0.5,.5}
\definecolor{dblue}{rgb}{.2,0.2,.8}
\definecolor{dgreen}{rgb}{.2,0.5,.2}

\usepackage{amsthm,amsmath,amssymb}
\usepackage{mathrsfs}

\graphicspath{{Figures/}}

\begin{document}

\title{Practical quantum simulation of small-scale non-Hermitian dynamics}

\author{Hongfeng Liu}
\affiliation{Shenzhen Institute for Quantum Science and Engineering and Department of Physics, Southern University of Science and Technology, Shenzhen, 518055, China}
\affiliation{International Quantum Academy, Shenzhen, 518055, China}
\affiliation{Guangdong Provincial Key Laboratory of Quantum Science and Engineering, Southern University of Science and Technology, Shenzhen, 518055, China}

\author{Xiaodong Yang}
\email{yangxd@sustech.edu.cn}
\affiliation{Shenzhen Institute for Quantum Science and Engineering and Department of Physics, Southern University of Science and Technology, Shenzhen, 518055, China}
\affiliation{International Quantum Academy, Shenzhen, 518055, China}
\affiliation{Guangdong Provincial Key Laboratory of Quantum Science and Engineering, Southern University of Science and Technology, Shenzhen, 518055, China}

\author{Kai Tang}
\affiliation{Shenzhen Institute for Quantum Science and Engineering and Department of Physics, Southern University of Science and Technology, Shenzhen, 518055, China}
\affiliation{International Quantum Academy, Shenzhen, 518055, China}
\affiliation{Guangdong Provincial Key Laboratory of Quantum Science and Engineering, Southern University of Science and Technology, Shenzhen, 518055, China}

\author{Liangyu Che}
\affiliation{Shenzhen Institute for Quantum Science and Engineering and Department of Physics, Southern University of Science and Technology, Shenzhen, 518055, China}
\affiliation{International Quantum Academy, Shenzhen, 518055, China}
\affiliation{Guangdong Provincial Key Laboratory of Quantum Science and Engineering, Southern University of Science and Technology, Shenzhen, 518055, China}

\author{Xinfang Nie}
\affiliation{Shenzhen Institute for Quantum Science and Engineering and Department of Physics, Southern University of Science and Technology, Shenzhen, 518055, China}
\affiliation{Guangdong Provincial Key Laboratory of Quantum Science and Engineering, Southern University of Science and Technology, Shenzhen, 518055, China}

\author{Tao Xin}
\affiliation{Shenzhen Institute for Quantum Science and Engineering and Department of Physics, Southern University of Science and Technology, Shenzhen, 518055, China}
\affiliation{International Quantum Academy, Shenzhen, 518055, China}
\affiliation{Guangdong Provincial Key Laboratory of Quantum Science and Engineering, Southern University of Science and Technology, Shenzhen, 518055, China}

\author{Jun Li}
\affiliation{Shenzhen Institute for Quantum Science and Engineering and Department of Physics, Southern University of Science and Technology, Shenzhen, 518055, China}
\affiliation{International Quantum Academy, Shenzhen, 518055, China}
\affiliation{Guangdong Provincial Key Laboratory of Quantum Science and Engineering, Southern University of Science and Technology, Shenzhen, 518055, China}

\author{Dawei Lu}
\email{ludw@sustech.edu.cn}
\affiliation{Shenzhen Institute for Quantum Science and Engineering and Department of Physics, Southern University of Science and Technology, Shenzhen, 518055, China}
\affiliation{International Quantum Academy, Shenzhen, 518055, China}
\affiliation{Guangdong Provincial Key Laboratory of Quantum Science and Engineering, Southern University of Science and Technology, Shenzhen, 518055, China}

\date{\today}

\begin{abstract}
Non-Hermitian quantum systems have recently attracted considerable attention due to their exotic properties.
Though many experimental realizations of non-Hermitian systems have been reported, the non-Hermiticity usually resorts to the hard-to-control environments and cannot last for too long times.
An alternative approach is to use quantum simulation with the closed system, whereas how to simulate non-Hermitian Hamiltonian dynamics remains a great challenge.
To tackle this problem, we propose a protocol which combines a dilation method with the variational quantum algorithm.
The dilation method is used to transform a non-Hermitian Hamiltonian into a Hermitian one through an exquisite quantum circuit,
while the variational quantum algorithm is for efficiently approximating the complex entangled gates in this circuit.
As a demonstration, we apply our protocol to simulate the dynamics of an Ising chain with nonlocal non-Hermitian perturbations, which is an important model to study quantum phase transition at nonzero temperatures.
The numerical simulation results are highly consistent with the theoretical predictions, revealing the effectiveness of our protocol.
The presented protocol paves the way for practically simulating small-scale non-Hermitian dynamics.
\end{abstract}

\maketitle
\section{INTRODUCTION}
Recent years have witnessed ongoing interests in exploring non-Hermitian phenomena \cite{el2018non,ashida2020non}.
Quantum systems driven by non-Hermitian Hamiltonians can lead to unconventional  and exclusive features, such as the non-Hermitian topological band \cite{ashida2020non,RevModPhys.93.015005}, the non-Hermitian skin effect \cite{PhysRevLett.121.086803}, the quantum critical phenomenon \cite{ashida2017parity}, chiral population transfer \cite{xu2016topological, doppler2016dynamically}, and anomalous bulk-boundary correspondence \cite{PhysRevLett.116.133903,PhysRevLett.121.026808,PhysRevLett.121.136802}.
However, experimentally investigating non-Hermitian physics is very challenging, particularly for achieving long-time dynamics in many-body systems \cite{roccati2022non}, because non-Hermiticity usually originates from particle loss and decoherence assigned by the environment, which is hard to manipulate \cite{li2019observation,ruter2010observation,miri2019exceptional,gao2015observation,zhang2017observation}.

In order to practically explore non-Hermitian physics, there have developed many strategies to simulate non-Hermitian Hamiltonian dynamics using closed quantum systems. By averaging the measured results over multiple noise configurations based on the stochastic Schr\"odinger equation, the non-Hermitian dynamics can be achieved,  yet it needs tremendous experimental resources thus is hard to scale
\cite{diosi1997non,gardiner2015quantum,lin2022experimental}. Researchers also suggest dilating a non-Hermitian Hamiltonian into a Hermitian one in a higher-dimensional Hilbert space, but they are only applicable for special types of parity-time-symmetric systems \cite{PhysRevLett.101.230404,PhysRevLett.119.190401,zheng2018duality,PhysRevLett.123.080404,PhysRevB.104.035153}. Remarkably, a recent work in Ref. \cite{wu2019observation} proposes a powerful method that can dilate a general non-Hermitian Hamiltonian into a Hermitian one, by executing a carefully designed quantum circuit on the ancilla-assisted system. Except for single-qubit rotations, this circuit contains a complex entangled unitary operation, which brings a new challenge for experimental realization, especially in the multiqubit case \cite{wu2019observation}. Conventional strategies for realizing a  unitary operation include matrix decomposition \cite{PhysRevA.52.3457,dogra2021quantum} and quantum simulation using the Suzuki-Trotter formulas \cite{trotter1959product,suzuki1976relationship,RevModPhys.86.153}. However, the former one is actually a highly nontrivial task, and the latter one may not have an available form for specific quantum systems. In addition, both of these two strategies inevitably produce exponentially growing operations, so the accumulated experimental errors will be serious. Therefore much more effort should be made on developing feasible methods that enable practical realization of the dynamics of non-Hermitian Hamiltonians.

\begin{figure*}
    \includegraphics[width=0.68\linewidth,height=0.5\linewidth]{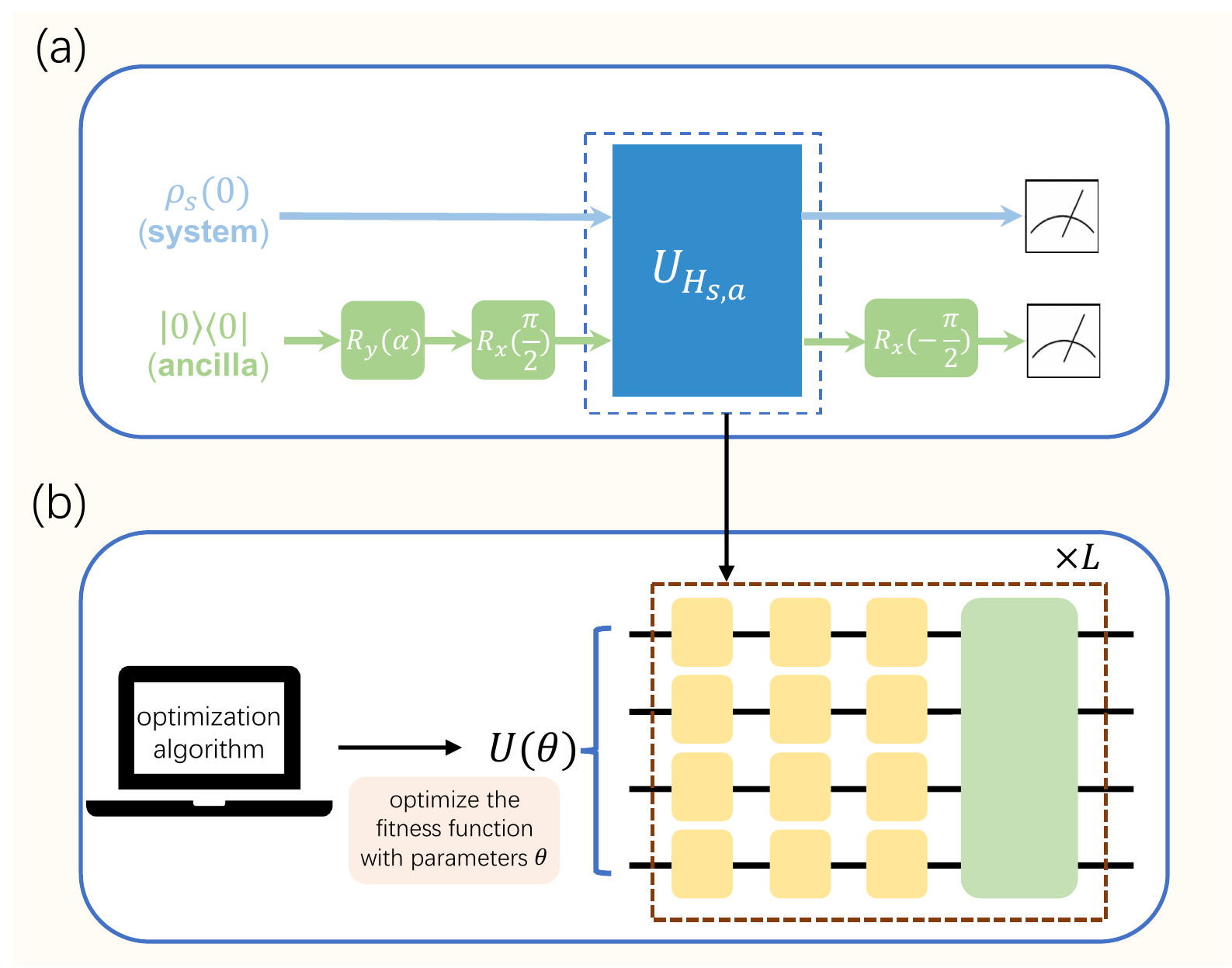}
    \caption{Schematic diagram of simulating the time evolution governed by a small-scale non-Hermitian Hamiltonian.
    (a) Quantum circuit for the dilation method. To simulate the dynamics of a non-Hermitian Hamiltonian $H_s(t)$, we introduce a dilated state $\rho_{s,a}(t)=\rho_s(t) \ket{-}\langle-|+\rho_s(t)\eta^{\dag}(t)\ket{-}\langle+|+\eta(t)\rho_s(t)\ket{+}\langle-|+\eta(t)\rho_s(t)\eta^{\dag}(t)\ket{+}\langle+|$ governed by a carefully designed dilated Hermitian Hamiltonian $H_{s,a}(t)$, where $\ket{-}=(\ket{0}-i\ket{1})/\sqrt{2}$ and $\ket{+}=-i(\ket{0}+i\ket{1})/\sqrt{2}$, and $\eta(t)$ is an appropriate linear operator. To realize this state, the system is initialized as an arbitrary state $\rho_s(0)$, and meanwhile the ancillary qubit is first prepared at $|0\rangle\langle 0|$ and then rotated by $R_y(\alpha)$ and $R_x(\pi/2)$ with  $\alpha=2\arctan\eta_0$. The joint system then undergoes the evolution driven by $H_{s,a}(t)$, i.e., $U_{H_{s,a}(t)}$. Afterwards, a rotation $R_x(-\pi/2)$ is applied on the ancilla. The non-Hermitian system's final state $\rho(t)$ can be extracted by measuring the subspace where the ancillary qubit is $|0\rangle\langle0|$.
  (b) Parameterized circuit $U(\theta)$ of the VQA.  We use the VQA to efficiently approximate the small-scale evolution operator $U_{H_{s,a}(t)}$. The VQA encodes parameters with $L$ layers of building blocks, with each block containing a limited number of single-qubit rotations (yellow square boxes) and an entangled gate (green square box). These parameters are then optimized with an appropriate optimization algorithm by setting a suitable {{fitness function}}.}
    \label{Fig1}
    \end{figure*}

In this work we propose to combine the dilation method \cite{wu2019observation} with the variational quantum algorithm (VQA) \cite{moll2018quantum,cerezo2021variational} for practically simulating small-scale non-Hermitian Hamiltonian dynamics.
VQA functions by iteratively optimizing a parameterized quantum circuit  with limited operations and circuit depth and thus has found plenty of applications in finding ground states and excited states for quantum chemistry \cite{kandala2017hardware,PhysRevX.6.031007,mccaskey2019quantum,PhysRevX.8.031022}. Moreover, it has also been used to approximate quantum processes and quantum gates \cite{PhysRevLett.125.010501,mcardle2019variational,heya2018variational,PhysRevResearch.3.033083}.
Here we bypass the difficulty of realizing the complex entangled operation in the dilatation method with the use of VQA.
This leads to an important advantage that this complex entangled operation can be realized in a much shorter time.
To optimize the parameters in VQA, it is routine to apply gradient-based algorithms \cite{PhysRevA.99.032331}.
However, as the system size and the circuit depth increase, the convergence speed of the conventional gradient-based algorithms  usually becomes very slow due to the quickly enlarged parameter space. To mitigate this issue, we utilize the backpropagation strategy, which is widely used in machine learning \cite{hecht1992theory}, to accelerate the optimization process.

To verify the effectiveness of our protocol, we apply it to the problem of simulating the evolution of a quantum Ising chain with nonlocal non-Hermitian perturbations \cite{PhysRevLett.126.116401}. This physical model is important for understanding the quantum phase transition (QPT) of spin systems at nonzero temperatures. However, its experimental realization faces great challenges mainly due to the  non-Hermitian terms. With our protocol we numerically simulate the evolution dynamics of this model up to five qubits.  We use the Loschmidt echo and its time average to characterize the QPT under  multiple conditions. The results are well consistent with the theoretically calculated results. Our protocol can be very helpful for experimentally realizing small-scale non-Hermitian Hamiltonian dynamics. 

The outline of this work is organized as follows. We first introduce the procedure of our protocol for simulating small-scale non-Hermitian Hamiltonian dynamics in Sec. \ref{framework}. Numerical simulations of applying our protocol to simulate the spin dynamics with a non-Hermitian perturbation are demonstrated in Sec. \ref{numerics}. Finally, discussions and outlook are presented in Sec. \ref{discussion}.

\section{Methodology}\label{framework}
We consider the task of simulating small-scale non-Hermitian Hamiltonian dynamics with closed quantum systems. The schematic diagram of our protocol is shown in Fig. \ref{Fig1}. Briefly speaking, our protocol consists of two parts: (1) dilate the non-Hermitian Hamiltonian into a Hermitian  one using the quantum circuit depicted in Fig. \ref{Fig1}(a), and (2) approximate the entangled operation in the above circuit with VQA; see Fig. \ref{Fig1}(b). In the following we introduce our protocol in detail.

\subsection{Procedures of the dilation method}
\begin{figure}
    \includegraphics[width=\linewidth]{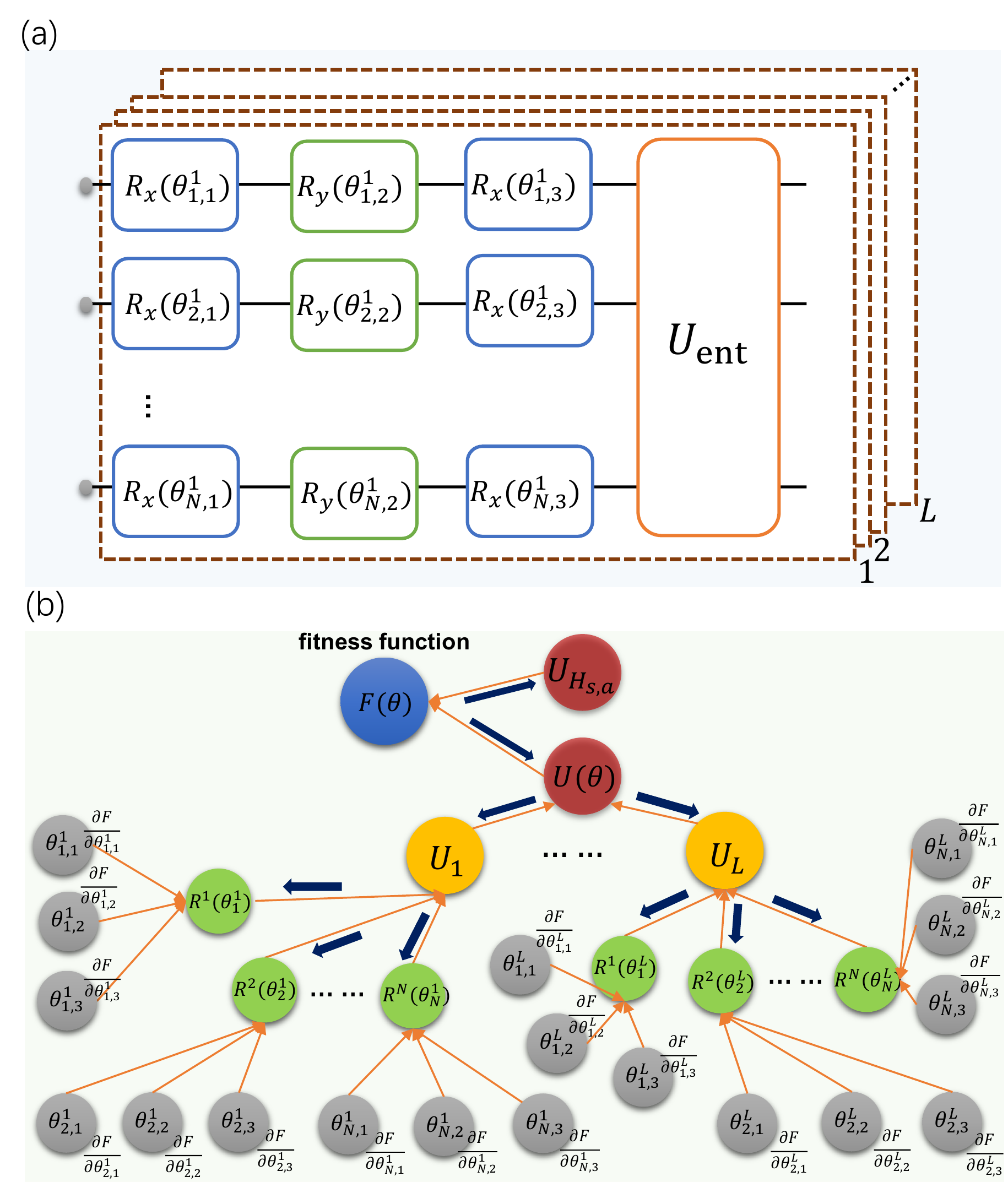}
    \caption{Building blocks of the VQA and schematic diagram of the backpropagation strategy.
   (a) VQA parameterizes a quantum circuit with $L$-layer building blocks, with each block containing finite single-qubit rotations {$R_\gamma(\theta_{\eta,k}^l)$} and a nonlocal gate $U_{\text{ent}}$, where {$\gamma=x,y; k=1,2,3; \eta=1,2,...,N; l=1,2,...,L$.}
  (b) {Gradients of the fitness function $F(\theta)$ with respect to $\theta_{\eta,k}^l$ are computed according to the illustrated chain rule. }{{The navy arrows guide the chain rule for calculating the gradients of $F(\theta)$, while the orange arrows illustrate the concept of  backpropagation.}} The backpropagation strategy suggests avoiding redundant calculations of the intermediate gradient nodes (e.g., yellow and green circles) for accelerating the convergence speed of the gradient-based algorithm.}
    \label{c and p}
    \end{figure}

We first describe the principle of dilating a non-Hermitian and generally time-dependent Hamiltonian into a Hermitian Hamiltonian with one single ancillary qubit \cite{wu2019observation}.
For such a system with Hamiltonian ${H}_s(t)$, its quantum state $\rho_s(t)$ satisfies the equation 
\begin{equation}\label{scheq}
	i\partial_t\rho_s(t)={H}_s(t)\rho_s(t)-\rho_s(t){H}_s^{\dag}(t).
\end{equation}
To realize this evolution in a quantum system, an ancillary qubit is introduced to dilate ${H}_s(t)$ into a Hermitian Hamiltonian $H_{s,a}(t)$, which is governed by the equation
 \begin{equation}\label{schsa}
 	i\partial_t\rho_{s,a}(t)=H_{s,a}(t)\rho_{s,a}(t)-\rho_{s,a}(t)H_{s,a}(t).
 \end{equation} 
Here $\rho_{s,a}(t)$ is the dilated state and can be expressed as
 \begin{align}\label{rsa}
 	 	\rho_{s,a}(t)
 	 	&=\rho_s(t) \ket{-}\langle-|+\rho_s(t)\eta^{\dag}(t)\ket{-}\langle+| \\\nonumber
 	 	&+\eta(t)\rho_s(t)\ket{+}\langle-|+\eta(t)\rho_s(t)\eta^{\dag}(t)\ket{+}\langle+|,
 \end{align}
where $\ket{-}=(\ket{0}-i\ket{1})/\sqrt{2}$ and $\ket{+}=-i(\ket{0}+i\ket{1})/\sqrt{2}$ are the eigenstates of the Pauli operator $\sigma_y$ with respect to the ancillary qubit, and $\eta(t)$ serves as a suitable linear operator for connecting the system Hamiltonian with the dilated Hamiltonian. 
  From Eq. (\ref{rsa}) we clearly see that the evolution of $\rho_s(t)$ can be obtained by measuring the state $\rho_{s,a}(t)$ in the subspace where the ancillary qubit is at state $\ket{-}\langle-|$. Now the key issue becomes how to construct the explicit form of $H_{s,a}(t)$. By expressing $H_{s,a}(t)$ in subspaces and substituting Eqs. (\ref{scheq}) and (\ref{rsa}) into Eq. (\ref{schsa}), we can get $i\partial_t M(t)=H_s^{\dagger}(t)M(t)-M(t)H_s(t)$ with $M(t)\equiv\eta^{\dag}(t)\eta(t)+I$, where $M(0)$ should be chosen to ensure that $M(t)-I$ keeps positive for all $t$, and $I$ represents the identity. Subsequently, we substitute the above relation into Eq. (\ref{schsa}) and observe that the choice of the dilated Hermitian Hamiltonian $H_{s,a}(t)$ is not unique. Therefore we can select 
\begin{equation}
H_{s,a}(t)=\varLambda(t)\otimes I+\varGamma(t)\otimes \sigma_z,
\label{Hsa}
\end{equation}
where $\varLambda(t)=\{{H_s(t)}+[i\frac{d}{dt}\eta(t)+\eta(t){H_s}(t)]\eta(t)\}M^{-1}(t)$ and $\varGamma(t)=i[{H_s}(t)\eta(t)-\eta(t){H_s}(t)-i\frac{d}{dt}\eta(t)]M^{-1}(t)$ with $\eta(0)=\eta_0 I, \eta_0=2$. Here $\sigma_z$ is the Pauli operator. In this way the dilated state can be calculated using $H_{s,a}(t)$, and the evolution of $\rho_s(t)$ is also available.

Concretely, the above described dilation method can be realized by the quantum circuit shown in Fig. \ref{Fig1}(a). The system's initial state is arbitrary and denoted as $\rho_s(0)$; meanwhile the ancillary qubit is initialized at $|0\rangle\langle0|$. Two rotations $R_y(\alpha)$ and $R_x(\pi/2)$ are then successively applied on the ancilla, where $R_\gamma(\beta)$ represents a $\beta$-angle rotation along the $\gamma$ axis and $\alpha=2\arctan\eta_0$. Afterwards, the system and the ancilla jointly undergo the evolution driven by $H_{s,a}(t)$. Finally, a rotation $R_x(-\pi/2)$ is applied on the ancillary qubit. With these operations, the total final state 
\begin{align}
		\rho_{s,a}(t)
		&=\rho_s(t) \ket{0}\langle0|+\rho_s(t)\eta^{\dag}(t)\ket{0}\langle1| \\ \nonumber
		&+\eta(t)\rho(t)\ket{1}\langle0|+\eta(t)\rho(t)\eta^{\dag}(t)\ket{1}\langle1|,
\end{align}
 thus $\rho_s(t)$ can be extracted by measuring in the subspace where the ancilla is $|0\rangle\langle0|$. However, in this quantum circuit, the evolution governed by $H_{s,a}(t)$ is a general {multiqubit} unitary operation, which brings a great challenge for experimental realizations. In the following we discuss how to solve this problem.

\subsection{VQA for approximating the evolution governed by $H_{s,a}(t)$}

To realize a general unitary operation, a direct way is to decompose it into basic operations, usually in terms of single-qubit gates and CNOT gates.  However, it has been proved that synthesizing any $N$-qubit $U(2^N)$ gate needs an order of $\emph{O}(4^N)$ basic operations \cite{PhysRevA.52.3457,shende2005synthesis}. This matrix decomposition task is highly complicated, and the exponentially increasing basic gates will introduce serious accumulated errors. An alternative way is to use quantum simulation with the Trotter-Suzuki formulas \cite{trotter1959product,suzuki1976relationship,RevModPhys.86.153}. Assume a general Hamiltonian that consists of many local-interaction terms, i.e., $H=\sum_l H_l$. The first step is to  break up the total evolution time $t$ into a sufficiently large number of equal slices, and the duration of each slice is denoted as $\Delta t$. Thus the evolution governed by $H$ can be calculated by $U=(e^{-i H \Delta t})^{t/\Delta t}$. As in most cases $[H_l,H_l'] \neq 0$, we then need to approximate $e^{-i H \Delta t}$ by the Trotter-Suzuki formulas. For example, the first-order Trotter-Suzuki formula leads to $U\approx(\prod_le^{-i H_l \Delta t})^{t/\Delta t}$. It is worth mentioning that the local term $H_l$ may not be available in specific quantum systems and thus more operations are needed to make a transformation. Therefore there will be a huge number of separated operations, especially when the higher-order Trotter-Suzuki formulas are applied for better precision, which can lead to very large errors.


As the above-mentioned two approaches face great challenges in efficiently realizing a general unitary gate, we thus need more practical methods.
The VQA is routinely used to find the eigenstates of molecules, but it can also be applied to many other problems, such as simulating the dynamics of quantum systems and solving linear  equations \cite{cerezo2021variational}. Here we utilize the VQA to approximate the evolution governed by $H_{s,a}(t)$, namely, $U_{H_{s,a}(t)}$. It is worth noting that similar tasks of studying how to decompose a known, complex unitary operator into an experiment-friendly circuit have been explored in many cases \cite{PhysRevLett.116.230504,PhysRevA.105.062421}. 
Usually the VQA parametrizes a quantum circuit $U(\theta)$ with $L$ sequentially applied layers (called ansatz), with each layer involving a limited number of single-qubit rotations and nonlocal gates. The specific structure of an ansatz is generally relevant to the problems at hand, but some ans\"atze can work even when no relevant information is readily known. Here we use such an ansatz called hardware-efficient ansatz, as shown in Fig. \ref{c and p}(a). The hardware-efficient ansatz determines the form of the nonlocal gates by the connectivity and interactions specific to a quantum hardware, which can reduce the circuit depth and thus has witnessed many practical applications \cite{kandala2017hardware,kokail2019self}. Concretely, the parameterized quantum circuit can be expressed as
\begin{equation}
{U(\theta)=\prod^L_{l=1}U_l,U_l= U_{\text{ent}} [\otimes^N_{\eta=1} R^\eta(\theta^l_{\eta})]},
\end{equation}
{where $R^\eta(\theta^l_{\eta})=R^\eta_x(\theta^l_{\eta,3}) R^\eta_y(\theta^l_{\eta,2}) R^\eta_x(\theta^l_{\eta,1})$, with $R^\eta_\gamma(\theta_{\eta,k}^l)$ representing $\theta_{\eta,k}^l$ rotation along the $\gamma$ axis applied on the $\eta$th qubit of the $l$th layer ($\gamma=x,y; k=1,2,3$), and $U_{\text{ent}}$ represents the hardware-specific nonlocal gate.}

On the other hand, as $H_{s,a}(t)$ is time dependent, we calculate the target evolution operator $U_{H_{s,a}(t)}$ by dividing the evolution time $t$ into sufficiently large $M$ equal segments, with the duration of each segment denoted as $\Delta t$. Thus we can get
\begin{equation}
	U_{H_{s,a}(t)}=\mathcal{T} e^{-i\int^{t}_{0}dtH_{s,a}(t)}=\prod_{m=1}^M e^{-i \Delta t H_{s,a}(m\Delta t)}.
\end{equation}
It is worth noting that we simply treat the evolution $U_{H_{s,a}(t)}$ as a whole so that the approximated operator $U(\theta)$ is very likely to be realized in a much shorter time. This is a great advantage for exploring the long-time behavior of the non-Hermitian Hamiltonian dynamics.


In the VQA, the parameterized circuit $U(\theta)$ needs to be optimized with appropriate optimization algorithms by setting a suitable fitness function. Here we set quantum gate fidelity as a fitness function, namely,
\begin{equation}\label{cost}
	F(\theta)=|\text{Tr}(U^\dag_{H_{s,a}(t)} U(\theta))|^2/4^N.
\end{equation}
Generally speaking, the optimization algorithms that have been used in the VQA can be classified into three categories, i.e., gradient-based algorithms \cite{PhysRevA.99.032331,PhysRevA.103.012405,PhysRevA.98.032309}, gradient-free algorithms \cite{zhu2019training}, and machine learning \cite{havlivcek2019supervised,PhysRevLett.128.120502}.
In our protocol we utilize the common gradient-based algorithm to accomplish the optimization.
However, as the system size becomes larger and the number of to-be-optimized parameters increases, the convergence speed of the gradient-based algorithm  can be significantly reduced.

To mitigate this issue, we combine the backpropagation strategy \cite{rumelhart1986learning}, which is widely used in machine learning, to accelerate the convergence process. {{The backpropagation strategy computes the gradients of the fitness function by the chain rule and then iterates backward from the last term to avoid redundant calculations of the intermediate terms in the chain rule \cite{werbos1990backpropagation}.}}
Specifically, the gradients of the {{fitness function}} in Eq. (\ref{cost}) with respect to the parameters $\theta^l_{n,k}$ can be calculated by the chain rule,
\begin{equation}
	\frac{\partial F}{\partial \theta^l_{n,k}}=\frac{\partial F}{\partial U}\frac{\partial U}{\partial U_l}\frac{\partial U_l}{\partial R^n}\frac{\partial R^n}{\partial \theta^l_{n,k}},
\end{equation}
as shown in Fig. \ref{c and p}(b). During the optimization, there exist many gradient nodes that are repeatedly calculated. For example, in each iteration the calculations of ${\partial F}/{\partial \theta^l_{n,k}}$ with the same $l$ and $n$ actually share the same gradient nodes ${\partial F}/{\partial U}$, $\partial U/\partial U_l$, and $\partial U_l/\partial R^n$. The backpropagation strategy suggests computing the above-shared gradient nodes only once and recording them for further use. This strategy can significantly save the computing time and thus accelerates the convergence.

\section{Numerical simulations}\label{numerics}
To demonstrate the effectiveness of our protocol, we apply it to simulate the evolution dynamics of an Ising chain with non-Hermitian perturbations \cite{PhysRevLett.126.116401}.
This physical model indicates that the QPT can be completely preserved at finite temperatures, and it presents an alternative approach for understanding the QPT of quantum spin systems at nonzero temperatures \cite{PhysRevLett.126.116401}.
However, it is a great challenge to realize such non-Hermitian Hamiltonians, especially for the multiqubit case. In the following we describe in detail how our protocol can accomplish this task.

\begin{figure*}
    \includegraphics[width=0.9\linewidth,height=0.44\linewidth]{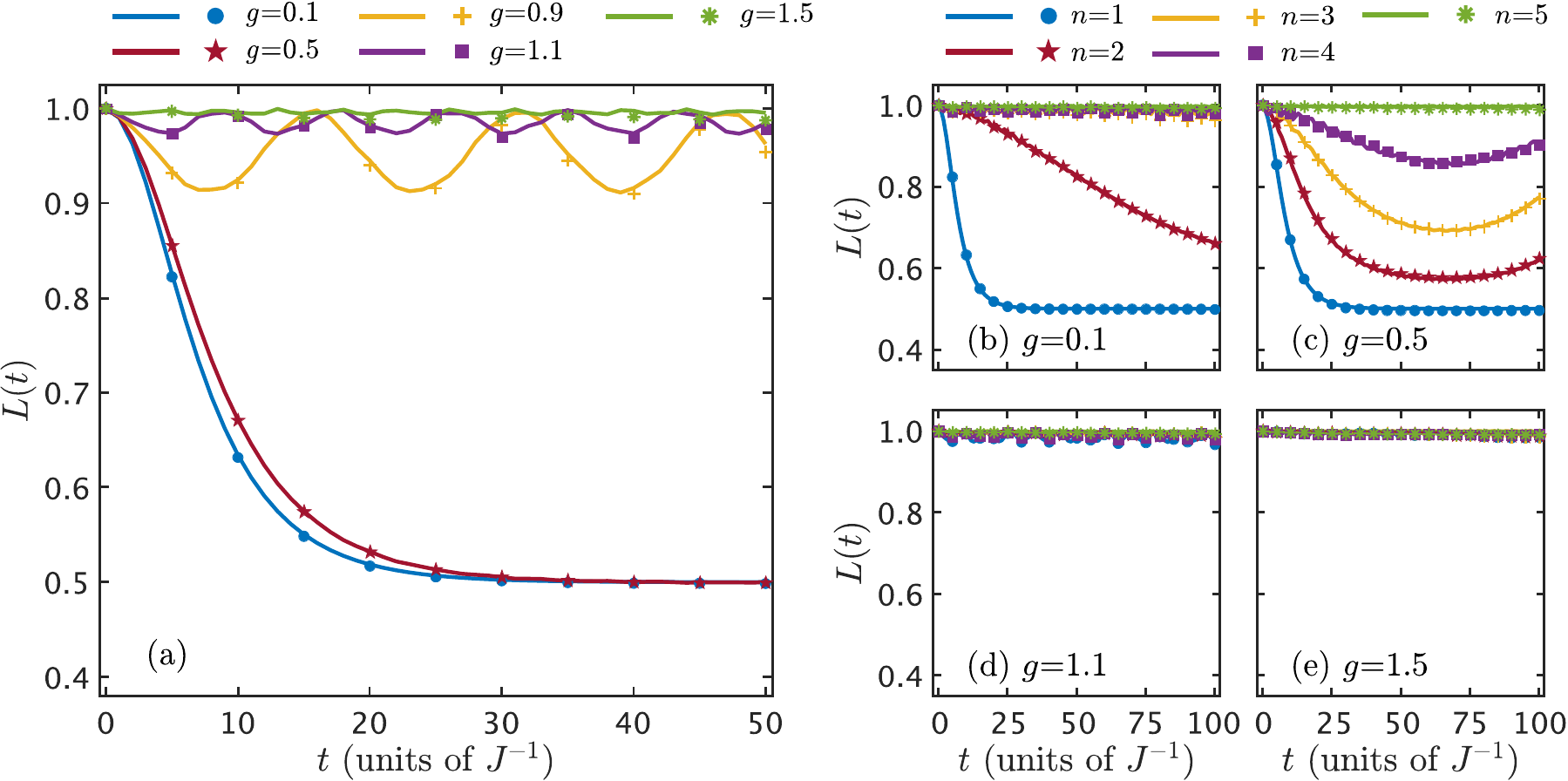}
    \caption{Comparison of the LEs with different Hamiltonian parameters. The solid lines and the markers represent the theoretical LEs [$L_\text{th}(t)$] and the simulated LEs [$L_\text{sim}(t)$], respectively.
    (a) Results for $H_s=H_0+\kappa D_1$ with different $g$.  The corresponding parameters are $N_s=5$, $J=1$, $\kappa=0.1$, and $\beta=10$. The profiles of the LEs in different phases are distinct,  converging to 1.0 for the paramagnetic phase and 0.5 for the ferromagnetic phase, respectively.
    (b-e) Results for $H_s=H_0+\kappa D_n$ ($n=1,2,3,4,5$) with different $g$. The corresponding parameters are $N_s=5$, $J=1$, $\kappa=0.1$, and $\beta=1$.
    The profiles of the LEs all converges to 1.0 for the paramagnetic phase when (d) $g=1.1$ and (e) $g=1.5$. However, for the case of the ferromagnetic phase (b) $g=0.1$ and (c) $g=0.5$, the LEs for large $n$ remain near 1.0, while the LEs for small $n$ gradually tend to 0.5.}
    \label{Lecho}
    \end{figure*}

\subsection{Model and methods}
{Consider an $N_s$-qubit Ising model under open boundary condition, which can be described by the following Hamiltonian:
\begin{equation}
	H_0=-J\sum^{N_s-1}_{n=1}\sigma_x^n \sigma_x^{n+1}+g\sum^{N_s}_{n=1}\sigma_z^n,
\end{equation}
where $\sigma_\gamma^n~(\gamma=x,y,z)$ are the Pauli operators applied on the $n$th site, $J$ represents the coupling strength, and $g~(g\geq 0)$ denotes the field strength. For simplicity, we assume $J=1$ in the following discussions.
In the thermodynamic limit we need to introduce a nonlocal non-Hermitian operator $D=\frac{1}{2}\sqrt{1-g^2}\sum^{N}_{n=1}g^{n-1}D_n$ \cite{PhysRevLett.126.116401}
with
\begin{equation}
	D_n=\prod_{l<n}(-\sigma_z^l)\sigma_x^n-i\prod_{l<N_s-n+1}(-\sigma_z^l)\sigma_y^{N_s-n+1}.
\end{equation} 
This perturbation term captures the impact of non zero temperature on the spin system.
Overall, the total system Hamiltonian can be defined as  
\begin{equation}\label{hsn}
H_s=H_0+\kappa D_n,
\end{equation}
where $\kappa$ is a real constant that satisfies $\kappa \ll g$. }

To simulate the evolution dynamics of the above non-Hermitian Hamiltonian using our protocol, we first use the method introduced in Sec. \ref{framework}.A to dilate it into an $N$-qubit Hermitian Hamiltonian with one ancillary qubit. {Thus we have the explicit relation $N=N_s+1$.} Next, in order to approximate $U_{H_{s,a}(t)}$ with the VQA, we need to focus on a specific quantum platform. Taking the nuclear magnetic resonance (NMR) system as an example, the nonlocal gate in this system can be conveniently expressed as
\begin{equation}
	U_{\text{ent}}=e^{-i H_{\text{int}} t_s}, H_{\text{int}}=\sum\nolimits_{i<j}^{N}\pi J_{i,j} \sigma_z^i\sigma_z^j/2,
\end{equation}
where $J_{i,j}$ is the coupling between the $i$th spin and the $j$th spin, and $t_s$ is the evolution time. As the NMR samples have fixed molecular structures, the couplings between the spins remain unchanged. Thus the above nonlocal gate is simply realized by letting the system undergo free evolution for a time period $t_s$, which is very favorable for the use of the  hardware-efficient ansatz; see Fig. \ref{c and p}(a). For an $N$-qubit parameterized quantum circuit with $L$ layers, we thus need $3NL$ single-qubit rotations and $N$ nonlocal gates. To optimize these rotation parameters, we apply the gradient-based algorithm with the backpropagation strategy \cite{paszke2017automatic}.

For exploring the QPT in the above physical model, we routinely introduce the Loschmidt echo (LE) \cite{PhysRevLett.96.140604}, i.e.,
\begin{equation}
L(t)=[\text{Tr}\sqrt{\sqrt{\rho_s(0)}\rho_s(t)\sqrt{\rho_s(0)}}]^2.
\end{equation}
The LE characterizes the degree of distinguishability between $\rho(0)$ and $\rho(t)$, which allows us to quantify the sensitivity of quantum evolution to perturbations. To study nonzero temperature QPT, we set the initial state as the thermal state  $\rho_s(0)=e^{-\beta H_0}/\text{Tr}e^{-\beta H_0}$ at temperature $\beta$ for the prequench Hamiltonian $H_0$. As the system state $\rho_s(t)$ obeys the equation $i {\partial \rho_s(t)}/ {\partial t}={H_s}\rho_s(t)-\rho_s(t){H_s}^{\dag}$ and $H_s$ is time independent, the system's final state with normalization can be directly calculated by $\rho_s(t)=e^{-i {H_s}t}\rho_s(0)e^{i{H_s^\dag}t}/\text{Tr}[e^{-i{H_s}t}\rho_s(0)e^{i{H_s^\dag}t}]$. In this way we can get the theoretical LE, marked as $L_{\text{th}}(t)$. On the other hand, when executing our protocol introduced in Sec. \ref{framework}, the system's final state obtained from the dilated circuit also needs to be normalized. This final state then leads to the simulated LE, marked as $L_\text{sim}(t)$.

\subsection{Simulation results}
We first consider the simulations for $H_s=H_0+\kappa D_1$, where $D_1$ is the dominant term of $D_n,n=1,2,...,N_s$. With $N_s=5$ qubits, we apply the above-introduced method to simulate its dynamical evolution. Specifically, we first substitute $H_s$ into Eq. (\ref{Hsa}), then construct the corresponding Hermitian Hamiltonian ${H_{s,a}}$, and finally simulate $U_{H_{s,a}}$ with VQA. {As for realizing $U_{\text{ent}}$ in the VQA with the NMR system, we choose the parameters of the seven-qubit sample $^{13}$C-labeled crotonic acid and set $t_s=0.0035~J^{-1}$ for the simulations; see the specific values $J_{ij}$ of this sample in Refs. \cite{PhysRevLett.118.150503,PhysRevLett.129.070502,PhysRevLett.129.100603}. } With sufficiently large number of layers $L=400$ and the initial guess $\theta_{n,k}^l=0$, we obtain the {fitness function} $F\geq 0.9995$. As described above, we use the LE to characterize the QPT in this physical model. The theoretical LEs and the simulated LEs are shown in  Fig. \ref{Lecho}(a). It can be seen that the LEs tend to different values for different $g$. Specifically, when $g>1$, the system is in the paramagnetic phase; the non-Hermitian perturbation does not substantially affect the dynamics, hence the LEs satisfy $ L(t)\approx L(0)=1.0$. On the contrary, the system is in the ferromagnetic phase when $g<1$, and the dominant non-Hermitian term $D_1$ leads to the exceptional point dynamics. The exceptional point makes the thermal state approach to its half component in the ferromagnetic phase \cite{PhysRevLett.126.116401}, and thus the LEs converge to 0.5. Generally speaking, the simulated LEs (markers) and the theoretical LEs (solid lines) are well matched, revealing that our protocol can almost exactly simulate the evolution dynamics of this non-Hermitian physical model. In addition, for each point in Fig. \ref{Lecho}(a), the operation time of the dilated circuit with and without resorting to the VQA are  $400t_s=1.4 J^{-1}$ and $5 J^{-1}$, respectively. This advantage makes our protocol more favorable for simulating the long-time behavior of the non-Hermitian Hamiltonian dynamics.

Next we consider the Hamiltonian $H_s=H_0+\kappa D_n$ with $n=1,2,3,4,5$. We set $N_s=5$ and use the same simulation parameters as above, i.e., $L=400, \theta_{n,k}^l=0$ and $t_s=0.0045~J^{-1}$. The fitness functions for all the cases satisfy $F\geq 0.9995$. The theoretical LEs and the simulated LEs are shown in  Figs. \ref{Lecho}(b)-\ref{Lecho}(e). It can be seen that the long-time behavior of the LEs when $g>1$ is similar to that in the above case, all converging to 1.0; see Figs. \ref{Lecho}(d) and \ref{Lecho}(e). For $g<1$, we find that the LEs converge to 0.5 when $n=1$, which is the same as the above case; see Figs. \ref{Lecho}(b) and \ref{Lecho}(c). However, the LEs with other position-dependent parameters $n=$2--5 have different features. Specifically, for $n=$3--5 in Fig. \ref{Lecho}(b) with $g=0.1$ and $n=$5 in Fig. \ref{Lecho}(c) with $g=0.5$, the LEs remain at 1.0. {{It is worth noting that the LEs remain at 1.0 when $n=5$ regardless of the value of $g$. The reason behind this is that $D_1$ is the dominant term of $D_n$; the non-Hermitian effects become smaller and smaller when $n$ increases \cite{PhysRevLett.126.116401}.}}
For other $n$ values in Figs. \ref{Lecho}(b) and \ref{Lecho}(c), it can be seen that the LEs gradually tend to the profile of $n=1$ as $n$ becomes smaller, indicating that the LEs decay more rapidly as $n$ approaches the boundary. In all the compared cases, we clearly find that the simulated LEs and the theoretical LEs are well consistent with each other, showing the effectiveness of our protocol on simulating the non-Hermitian Hamiltonian dynamics again.

\begin{figure}
    \includegraphics[width=0.95\linewidth]{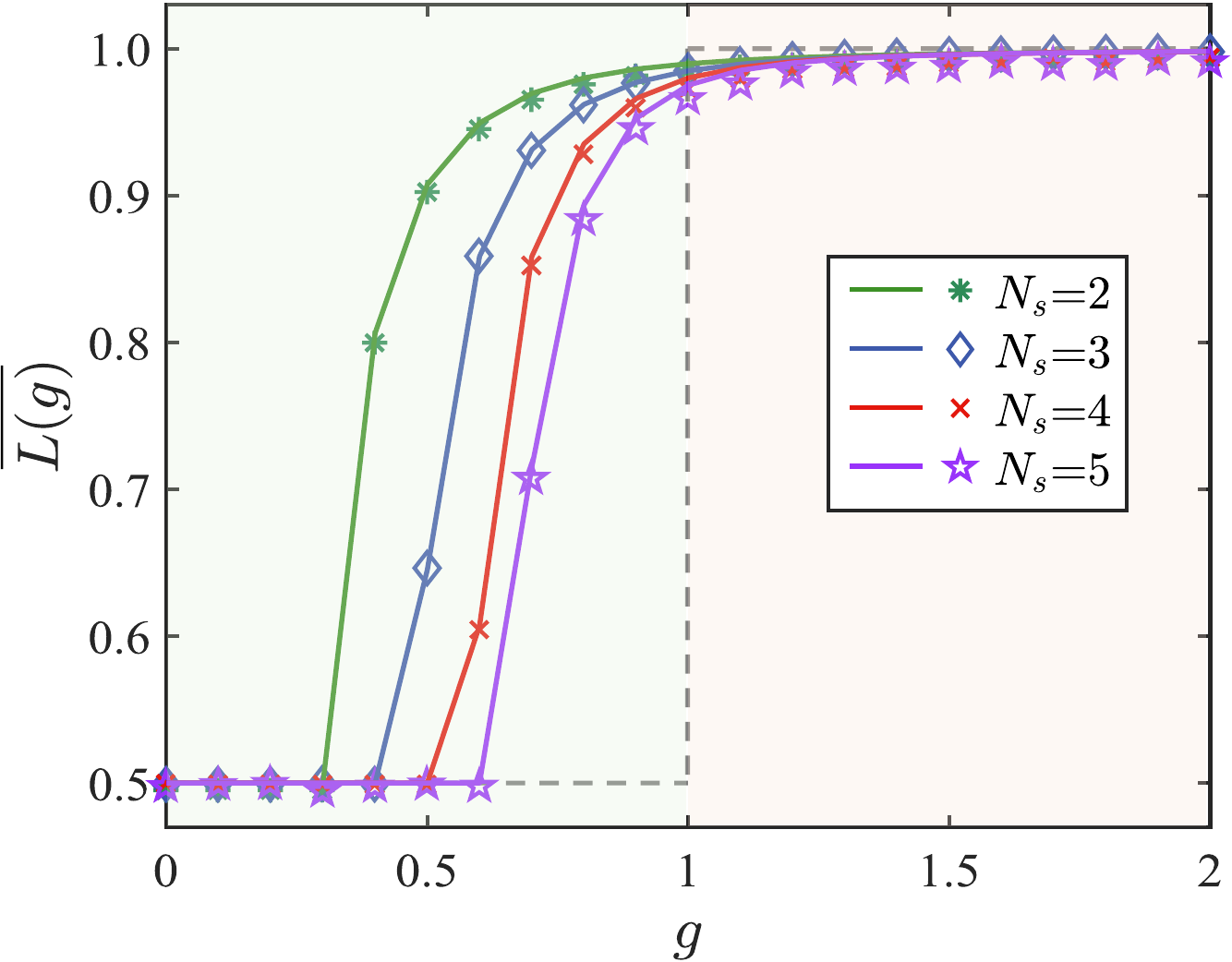}
    \caption{Average LEs as a function of $g$ for different $N_s$. The solid lines represent the theoretical average LEs ($\bar L_{\text{th}}$), while the markers are the simulated average LEs ($\bar L_{\text{sim}}$). The dashed line is the ideal average LEs for large $N_s$ limits, which separates the ferromagnetic phase ($g<1$, green background region) and the paramagnetic phase ($g>1$, pink background region). {{Here we set $\kappa=0.1, J=1, \beta=10 $, $\tau=500, T=500$, and $n=1$.}} It reveals that the average LEs tend to the prediction in the thermodynamic limit as $N_s$ increases.}
    \label{L_aver}
    \end{figure}

Finally, we check the average LE to determine the effect of $g$ with $g$ varying from 0 to 2. The average LE in the time interval [$\tau$, $\tau+T$] can be defined as $\bar{L}=\frac{1}{T}\int_{\tau}^{\tau+T} L(t)dt$, where $\tau \gg 1$. Similarly, we further mark the theoretical average  LE and the simulated average LE as $\bar L_{\text{th}}$ and $\bar L_{\text{sim}}$, respectively. For simplicity, we restrict ourselves to the Hamiltonian $H_s=H_0+\kappa D_1$ with $\kappa=0.1$. This is because $D_1$ is the dominant non-Hermitian term of $D_n$, and the impacts of other terms are very small; thus we can omit them. With the same simulation procedure and the same parameters, we obtain the results shown in  Fig. \ref{L_aver}. It can be seen that when $g$ is very small, the average LE remains at 0.5. As $g$ increases, the average LE for different $N_s$ rapidly rises to 1.0 and has a trend to gradually approach the thermodynamic limit (dashed line, $N_s$ is very large) as $N_s$ increases. Moreover, when $g>1$, the LEs remain unchanged at around 1.0. These results indicate that the LEs are also effective to  characterize the QPT at nonzero temperatures, even for small-scale systems. We clearly see that the simulated average LEs are well matched with the theoretical average LEs for all the tested $N_s$ and $g$, demonstrating the effectiveness of our protocol in simulating the evolution dynamics of non-Hermitian Hamiltonians.

\subsection{Computational cost analysis}

\begin{figure}
    \includegraphics[width=0.9\linewidth]{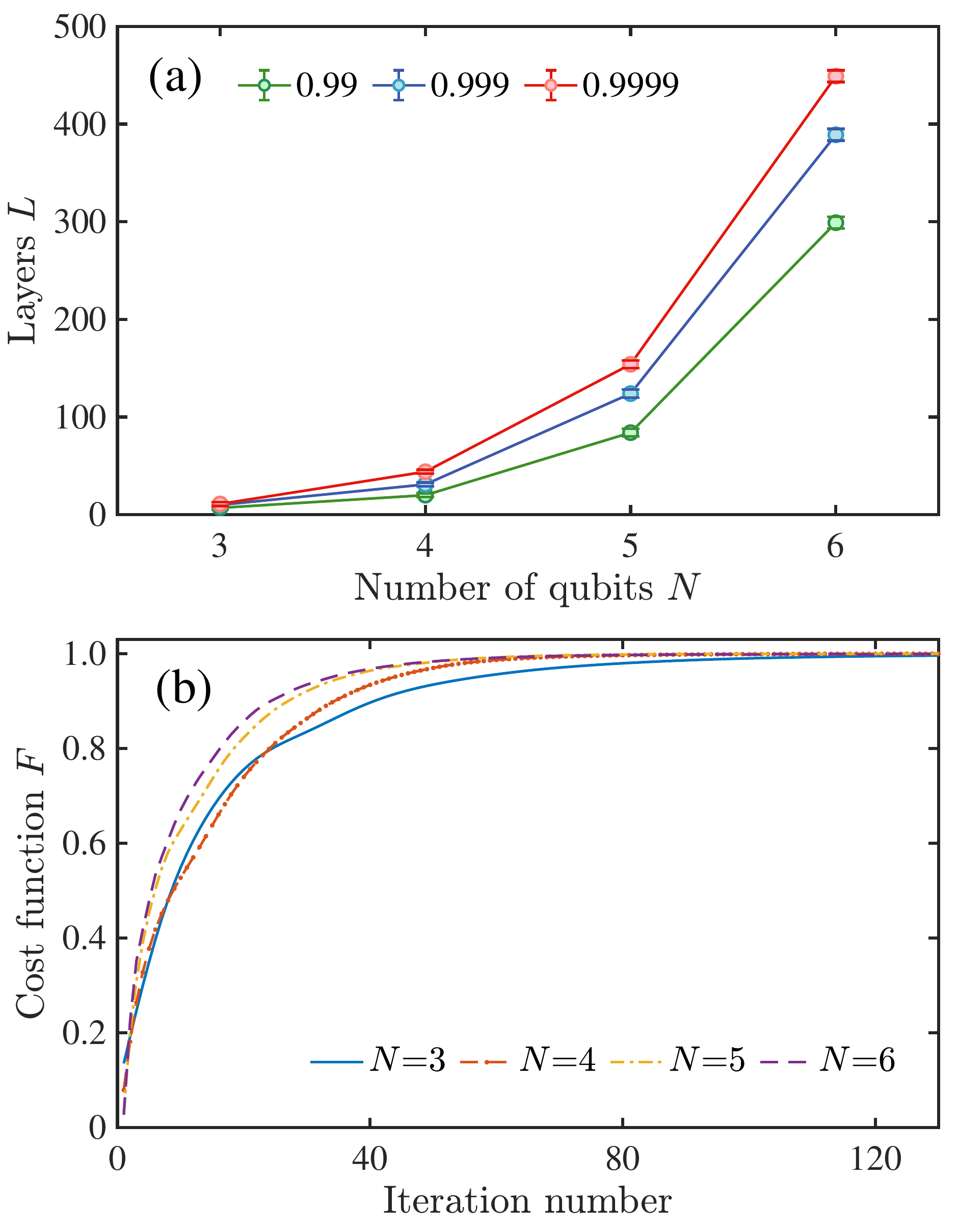}
    \caption{
    (a) Number of circuit layers $L$ needed for reaching a certain value of the {{fitness function}} $F$ vs the number of qubits $N$. The error bars are plotted with five runs using random initial parameters $\theta_{n,k}^l$.
    (b) {{Fitness function}} $F$ vs the iteration number for different numbers of qubits $N$.
}
    \label{Different layer}
    \end{figure}

As the VQA is a key part of our protocol, we here analyze its computational cost. In the above simulations, we set a sufficiently large number of layers $L=400$ to make sure that the approximated $U(\theta)$ is as close to the target gate $U_{H_{s,a}(t)}$ as possible, resulting in $F \geq 0.9995$ for all cases. However, the actual applications may not need such high values of the fitness function, and thus we can reduce the number of layers needed. This reduction is beneficial to the experimental applications, as real quantum platforms have finite coherence times which limits the number of quantum operations that can be faithfully realized. In addition, the VQA parameters can also be significantly reduced, which can accelerate the optimization process.
Here we explore the number of circuit layers required to achieve distinct values of the fitness function $F$ for different number of qubits $N$. The simulation results are shown in Fig. \ref{Different layer}(a). It can be seen that the number of layers can be reduced about 9\%--30\% or 33\%--55\%, if decreasing $F=0.9999$ to $0.999$ or $0.99$, respectively.  Therefore, we need to make a balance between the accuracy of the fitness function and the circuit layers a quantum platform can afford for specific applications. For NMR system, the decoherence time of the used sample is greater than 2 s \cite{knill2000algorithmic}, and the fidelities of single-qubit gates can approach 0.9999; thus the simulations here are roughly within the reach of current NMR techniques. 

We also explore the convergence speeds of the gradient-based algorithm with the backpropagation strategy for finding the optimal VQA parameters. Here we set a sufficiently large number of circuit layers $L$ for each number of qubits $N$ to guarantee that the optimization algorithm can converge in an acceptable time; see  the results in Fig. \ref{Different layer}(b). It can be seen that the {{fitness function}} $F$ for different numbers of qubits $N$ quickly converges within 120 iterations, reaching $F\geq0.995$ for all cases. This reveals that the backpropagation-enhanced optimization algorithm has a very fast convergence speed, and the increasing number of system dimensions and the VQA parameters do not influence the convergence speeds very much. This is an important feature of our protocol for efficiently simulating multiqubit non-Hermitian dynamics.

\section{Discussions and outlook}\label{discussion}
In summary, we propose a practical protocol for efficiently simulating the small-scale non-Hermitian Hamiltonian dynamics. The basic idea is to first dilate the non-Hermitian Hamiltonian into a Hermitian one with a carefully designed quantum circuit and then use the VQA with the hardware-efficient ansatz to approximate the complex entangled operation in this circuit. To optimize the VQA parameters, we apply the gradient-based optimization algorithm enhanced by the backpropagation strategy.
To show the effectiveness of our protocol, we use it to simulate the dynamics of an Ising chain with nonlocal non-Hermitian perturbations. With simulations up to five qubits, we demonstrate that the simulated results using our protocol are well matched with the theoretical predictions.
We hope the proposed protocol can soon find other applications and experimental verifications on various quantum platforms.

There are several aspects that are worth further efforts. First, for the specific problems at hand, problem-inspired ans\"atze \cite{cerezo2021variational} instead of the hardware-efficient ansatz may be more efficient to approximate the target unitary gate, which can significantly decrease the optimization complexity. Therefore suitable ans\"atze need to be chosen according to the problems and features of the available physical platform.
Second, there may exist many sources of noise in realizing the optimized VQA, so it is helpful to explore the behavior of our protocol when including typical noises in optimizing the VQA parameters \cite{wang2021noise,PhysRevA.104.022403}.
Third, the present method is restricted to small-scale systems (up to around 12 qubits) due to the difficulty of computing the general time-evolution operators on a classical computer \cite{lu2017enhancing}. However, as we are now in the noisy intermediate-scale quantum era \cite{preskill2018quantum}, considering the problem of simulating the non-Hermitian Hamiltonian dynamics up to tens of qubits will be highly interesting and demanded. This suggests that we combine more efficient time-evolution simulation strategies \cite{PhysRevA.103.023107}, more effective VQA ans\"atze and more powerful optimization algorithms to accomplish this task.

\begin{acknowledgments}
This work is supported by the National Key Research and Development Program of China (Grant No. 2019YFA0308100), the National Natural Science Foundation of China (Grants No. 12204230, No. 12104213, No. 12075110, No. 11825401, No. 11975117, No. 11905099, No. 11875159  and No. U1801661), the Guangdong Basic and Applied Basic Research Foundation (Grants No. 2019A1515011383 and No. 2020A1515110987), National Natural Science Foundation of China (12104213), Guangdong Basic and Applied Basic Research Foundation (2020A1515110987), the Guangdong International Collaboration Program (Grant No. 2020A0505100001), the Science, Technology and Innovation Commission of Shenzhen Municipality (Grants No. ZDSYS20170303165926217, No. KQTD20190929173815000, No. JCYJ20200109140803865, No. JCYJ20170412152620376 and No. JCYJ20180302174036418), the Open Project of Shenzhen Institute of Quantum Science and Engineering (Grant No. SIQSE202003), the Pengcheng Scholars, the Guangdong Innovative and Entrepreneurial Research Team Program (Grant No. 2019ZT08C044), and the Guangdong Provincial Key Laboratory (Grant No. 2019B121203002).
\end{acknowledgments}

%

%

\end{document}